\documentclass[reprint,amsmath,amssymb,aps,prl
]{revtex4-1}
\usepackage{graphicx}
\usepackage{dcolumn}
\usepackage{bm}
\usepackage{xcolor}
\usepackage{soul}
\usepackage{placeins}
\usepackage{hyperref}
\newcommand{\Duke}{Duke Quantum Center, Department of Electrical and Computer Engineering and Department of Physics \\ Duke University, Durham, NC 27708}
\begin{document}
\preprint{APS/123-QED}

\title{Fast programmable entanglement of Barium ion qubits using Rydberg states and AC-Stark shifts}

\author{A.R.~Vernon}
\email{adamvernon@proton.me}
\affiliation{\Duke}
\author{M.G. Peaks}
\affiliation{\Duke}

\date{\today}

\begin{abstract}
A scheme for excitation and individual addressing using Rydberg states of trapped Barium ions is presented for the purpose of fast gates and entanglement. Dipole matrix elements, dynamic polarizabilities, and one- and two-photon transition strengths are computed with a Supersymmetric Wentzel–Kramers–Brillouin (SWKB) method. A favorable two-photon excitation transition is identified, linking the 7s$_{1/2}$ state to high-lying Rydberg states, with the strongest transition found to be 7s$_{1/2}$ $\rightarrow$ 38s$_{1/2}$. Additionally, the 7s$_{1/2}$ state exhibits high polarizability around the telecommunications band at 1310 nm, enabling significant AC-Stark shift control with a turnkey laser at low power. This facilitates an individual addressing scheme by varying light intensity across an ion crystal, supporting sub-microsecond entangling gates between ion pairs with the strong and microwave-tunable Rydberg dipolar interaction. Selective addressing of individual ions by laser frequency tuning via the 6p$_{3/2}$ $\rightarrow$ 7s$_{1/2}$ transition is proposed.
\end{abstract}

\maketitle
\section{Introduction}
There is a strong need for quantum computer architectures to scale to thousands of qubits to achieve advantage over classical computers for many applications \cite{Monroe2013, Farrell2024}.
While Wigner ion crystals with hundreds of ions have been prepared and controlled, and scalable cooling to their ground states has been achieved \cite{Kiesenhofer2023}, ion-trap-based quantum computing still faces multiple scalability challenges. This work describes a novel scheme to scale quantum computers using the trapped Ba$^+$ atomic ion platform.

The Mølmer-Sørensen gate\cite{Sorensen1999}, a key two-qubit gate in ion-trap systems, has an entanglement time that scales linearly with the number of trapped ions\cite{Taylor2017}, resulting in gate times of hundreds of microseconds. The Mølmer-Sørensen gate has the additional advantage over the previously used Cirac-Zoller gate that the ions are not required to be in their motional ground state\cite{Cirac1995}.
Additionally, Raman transition lasers are increasingly used for qubit operations on ion chains. This requires laser power and the number of addressing beams to scale linearly with ion count to maintain individual addressability, while also facing challenges from crosstalk as the ion number grows. These factors are implemented most effectively with ion crystals configured in a 1D chain \cite{Moses2023}.

We propose a method to address scalability challenges by minimizing the positional dependence of ions, which additionally demands numerous DC and/or RF potential-shaping electrodes, introducing parasitic stray electric fields that degrade the high fidelity of trapped-ion operations\cite{An2022}. A recently demonstrated platform leverages Rydberg states of ions to achieve sub-microsecond entangling gates, utilizing their strong, tunable dipolar interactions\cite{Muller2008, Mokhberi2020}. 
Entanglement with Rydberg states exhibits low residual coupling to motional modes, contributing less than 10$^{-4}$ to gate error\cite{Zhang2020}; offering a key advantage for scaling entangling gates in large, n-ion crystals. Moreover, the Rydberg interaction can be implemented to create a `controlled phase gate'\cite{Rao2014, Zhang2020}, replacing the need for Raman lasers for two-qubit gates.

A major limitation of Rydberg architectures is the use of UV light to excite Rydberg states to maintain coherence, recently demonstrated in the Sr$^+$ system \cite{Zhang2020}. However, this entangling gate scheme is less dependent on precise laser positioning or ion shuttling with surface trap electrodes \cite{Moses2023}. An all-to-all connected architecture using Rydberg states has been proposed \cite{Bao2025}. Rydberg-based entanglement gates can operate on nanosecond timescales \cite{Chew2022}, comparable to super-conducting qubit architectures, which require greater engineering overhead to achieve gate fidelities close to those already standard in trapped-ion systems \cite{An2022}. This is due to the lower coherence times and high sensitivity of superconducting qubits to environmental noise\cite{Ithier2005}.

Combining the gate speed of Rydberg interactions with the fidelity of trapped ions could advance several fields, for example the neutral atom community has recently adopted microwave-dressing for tunable interaction strengths \cite{Giudici2024, Kurdak2025}, a technique which was required and developed for trapped Rydberg ion operations\cite{Zhang2020}.

To quantify the proposed Rydberg ion excitation addressing scheme for the Barium ion, dipole transition strengths were calculated. Precision atomic relativistic configuration-interaction\cite{Mitroy2010, Stroberg2019} or coupled-cluster\cite{Sahoo2008, Hagen2014} methods are typically used herein when available. However, they are often impractical to automate for high-lying states, like Rydberg states, due to large valence spaces. In this work, the Supersymmetric Wentzel–Kramers–Brillouin (SWKB) method was employed to compute transition dipole moments, one- and two-photon transition rates (to avoid UV transitions), and dynamic (laser frequency dependent) state polarizabilities for high-lying states, including relevant Rydberg states.

\section{Methods: Rydberg matrix dipole element calculation} \label{Sect:dipole}

\begin{figure}[t]
\includegraphics[width=8cm]{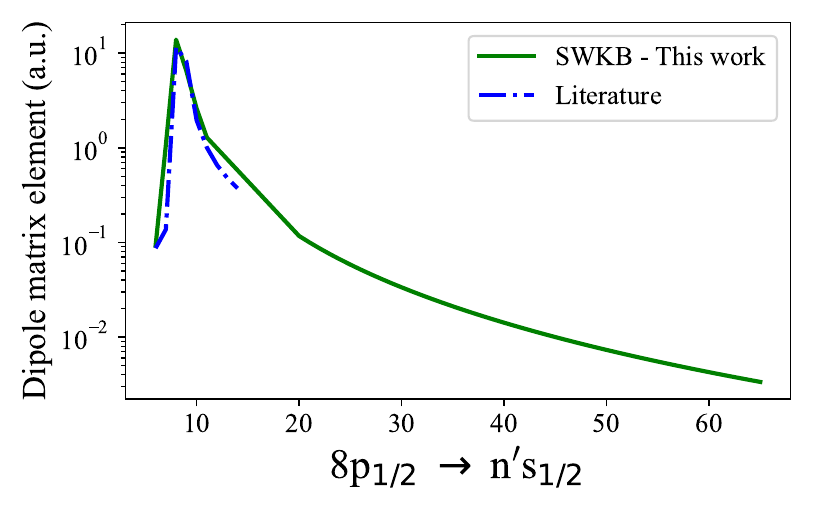}
\caption{\label{fig_dipole} Calculated SWKB dipole matrix elements from 8p$_{1/2}$ in Ba$^+$ to the (n$^\prime$)s$_{1/2}$ series, including Rydberg states. Literature values taken from the database of Ref.~\cite{Barakhshan2022}.}
\end{figure}

The general radial dipole moment matrix element is given by
\begin{equation}
R_{n l}^{n^{\prime} l^{\prime}}=\int r \psi_{n^{\prime} l^{\prime}}(r) \psi_{n l}(r) \mathrm{d} r, \quad l^{\prime}=l \pm 1  ,  \label{eqn_dipole}
\end{equation}

here n and l represent the principal and orbital quantum numbers, respectively, while $\psi_{n l}(r)$ is the reduced radial dipole wavefunction.
Naccache\cite{Naccache1972} suggested that semi-classical matrix elements can be enhanced by optimizing intermediate state parameters for a given intermediate state with energy E$_c$, a concept later validated by multiple studies\cite{Davydkin1981, Pankratov1992}. The SWKB method employed here is considered more accurate than SKB approaches \cite{Kamta1998}, as, for example, it does not require first-order Langer modifications for precise results \cite{Adhikari1988}.
This leads to the analytic expression of

\begin{equation}
\begin{aligned}
R_{n l}^{n^{\prime} l^{\prime}}=(-1)^{\Delta n+m} \frac{\mathrm{~d} a_{\mathrm{c}}}{\gamma} \{ & \left(\left(1-\varepsilon_{\mathrm{c}}\right) \gamma-\frac{\eta_{\mathrm{c}} \Delta_0}{\varepsilon_{\mathrm{c}}}\right) \frac{\sin \pi \gamma}{\pi \gamma} \\
& \left.+\frac{\eta_{\mathrm{c}} \Delta_0}{\varepsilon_{\mathrm{c}}} J_\gamma\left(-\gamma \varepsilon_{\mathrm{c}}\right)-J_\gamma^{\prime}\left(-\gamma \varepsilon_{\mathrm{c}}\right)\right\},
\end{aligned}
\label{eqn_Bnl}
\end{equation}

when taking the \textit{Bosonic} dipole matrix element from the SWKB derivation of Lagmago Kamta et al.\cite{Kamta1998}.
The main difficulty is in evaluating the Anger functions, $J_\gamma\left(-\gamma \varepsilon_{\mathrm{c}}\right)$ and their first derivatives $J_\gamma^{\prime}\left(-\gamma \varepsilon_{\mathrm{c}}\right)$, a simplification over previous WKB derivations that also involve evaluation of Weber functions \cite{Pankratov1992}. 
The Anger function is defined as  

\begin{equation}
    \mathbf{J}_\nu(z)=\frac{1}{\pi} \int_0^\pi \cos (\nu \theta-z \sin \theta) \mathrm{d} \theta, 
\end{equation}

using the generalized notation of $\nu$ for $\gamma$ and $z$ for $-\gamma \varepsilon_{\mathrm{c}}$. This can be evaluated by 

\begin{equation}
    \mathbf{J}_\nu(z)=\cos \left(\frac{1}{2} \pi \nu\right) S_1(\nu, z)+\sin \left(\frac{1}{2} \pi \nu\right) S_2(\nu, z),
\end{equation}

where sums $S_1(\nu, z)$ and $S_2(\nu, z)$ are

\begin{equation}
    S_1(\nu, z)=\sum_{k=0}^{\infty} \frac{(-1)^k\left(\frac{1}{2} z\right)^{2 k}}{\Gamma\left(k+\frac{1}{2} \nu+1\right) \Gamma\left(k-\frac{1}{2} \nu+1\right)},
\end{equation}

and

\begin{equation}
    S_2(\nu, z)=\sum_{k=0}^{\infty} \frac{(-1)^k\left(\frac{1}{2} z\right)^{2 k+1}}{\Gamma\left(k+\frac{1}{2} \nu+\frac{3}{2}\right) \Gamma\left(k-\frac{1}{2} \nu+\frac{3}{2}\right)},
\end{equation}

here $\Gamma$ is Euler's gamma function\cite{Abramowitz2012}.

The first derivative of $\mathbf{J}_\nu(z)$ was derived in Ref.\cite{Gaunt2017} to be given by

\begin{equation}
    \mathbf{J}_\nu(z)^{\prime}=\frac{\partial}{\partial z}\left(\mathbf{J}_v(z)\right)=\frac{1}{2}\left(\mathbf{J}_{v-1}(z)-\mathbf{J}_{v+1}(z)\right) .
    \label{eqn_JDer}
\end{equation}

Expressions \ref{eqn_Bnl} and \ref{eqn_JDer} are used herein to estimate the magnitudes between relatively high quantum number ($n\gtrsim$ 10) states. 

Energy levels are known for several Rydberg states of Ba$^{+}$\cite{Jones1989,Seng1998}, these are used to derive the quantum defect to account for the effect of the Ba$^{2+}$ core in these calculations.

The remaining terms of Eqn.~\ref{eqn_Bnl} are defined herein, where atomic units are used throughout:

$\Delta n$ is the difference in principle quantum numbers, $\Delta n = n'-n$, $m$ is the nearest integer (denoted by $\langle$ $\rangle$) in the expression 

\begin{equation}
    m=\left\langle\frac{\omega Z}{\left|E^{\prime}+E\right|^{3 / 2}}-s\right\rangle ,
\end{equation}

with the energies of the principle quantum number states $E$, and $\omega$ is the frequency difference between states.
The quantum defects, $\delta_n$ from literature\cite{Jones1989,Seng1998} are used to calculate the core-adjusted quantum number difference $s=\nu'-\nu$ where $\nu=n-\delta_n$. $Z$ is the core charge (here $Z=2$).

The terms $d$, $a_c$ and $\eta_c$ are defined in the following way:

\begin{equation}
    d=\frac{v_{\mathrm{c}}^3}{\bar{v}^3}, \quad a_{\mathrm{c}}=\frac{v_{\mathrm{c}}^2}{Z}, \quad \eta_{\mathrm{c}}=\frac{\lambda_{\mathrm{c}}}{v_{\mathrm{c}}}
\end{equation}

where

\begin{equation}
    v_{\mathrm{c}}=\sqrt[3]{\frac{\gamma Z^2 }{\omega}}, \quad \bar{\nu}=\frac{\nu'+\nu}{2}, \quad \lambda_c=\frac{l'+l+1}{2},
\end{equation}

$l$ are the orbital quantum numbers of the states and 

\begin{equation}
    \Delta_0=l'-l, \quad \gamma=m+s,\quad \mathrm{and}\quad \varepsilon_{\mathrm{c}}=\sqrt{1-\frac{\lambda_{\mathrm{c}}^2}{v_{\mathrm{c}}^2}}.
\end{equation}

The results of the calculations compared to the precision database of Ref.~\cite{Barakhshan2022} are shown in Fig.~\ref{fig_dipole} for the transition 8p$_{1/2}$ $\rightarrow$ ($n'$)s$_{1/2}$. Our SWKB values were used in this work where existing calculations were not available. These values were critical to assess the feasibility of the scheme described in this work. Additionally these one-photon transition strengths are shown in Table. ~\ref{tab_As}.

The subsequent section describes the Rydberg excitation process. To preserve coherence with ground state qubits, STImulated Rapid Adiabatic Passage (STIRAP) has been shown as an effective technique for exciting and interacting with Rydberg states \cite{Higgins2017, Zhang2020, Rao2014}.

\section{Methods: Two-photon transitions}

To eliminate the need for UV-light transitions, which are commonly used for Rydberg state excitation in ions, but can be destructive to surface traps and require specialized optics \cite{Mokhberi2020}, this work recommends employing two-photon transitions. The relatively low ionization potential of Barium compared to other alkali-like ions \cite{Karlsson1999}, along with conveniently spaced intermediate levels, enables strong two-photon transitions using light around 526 nm. This light is $\sim$118 GHz detuned from resonance with both one-photon transitions, facilitating excitation to the (n$^\prime$)s$_{1/2}$ Rydberg states which would otherwise be achieved via the equivalent two single-photon transitions: 7s$_{1/2}$ $\rightarrow$ 8p$_{1/2}$, followed by 8p$_{1/2}$ $\rightarrow$ (n$^\prime$)s$_{1/2}$.

The near-resonant condition is modeled in \cite{Grynberg1977}, given as

\begin{equation}
\begin{aligned}
&\mathrm{P}_{\mathrm{lR}}{ }^{2\gamma}(\mathrm{res})=\frac{1}{\Gamma_{\mathrm{R}}}\left(\frac{3}{\pi} \frac{r_0 \lambda_{\mathrm{l} \mathrm{i}} \lambda_{\mathrm{iR}}}{\hbar c}\right)^2\left(\frac{P}{A}\right)^2 \frac{\omega_{\mathrm{li} } \omega_{ \mathrm{iR}}}{\Delta \omega_{\mathrm{i} }^2} f_{\mathrm{li}} f_{\mathrm{iR}}  \\
& \times\left|\left\langle J_{\mathrm{R}} 1 m_{\mathrm{R}}-q \mid J_\mathrm{i} m_\mathrm{i}\right\rangle\left\langle J_\mathrm{i} 1 m_\mathrm{i}-q \mid J_{\mathrm{l}} m_{\mathrm{l}}\right\rangle\right|^2
\end{aligned}
\label{eqn_twop}
\end{equation}

Here, we apply this model to compare established literature values for the transition strengths, $A_{ij}$, of the one-photon transitions commonly used in quantum computing with Ba$^+$.
In Eqn.~\ref{eqn_twop} the lower 7s$_{1/2}$, intermediate 8p$_{1/2}$ states and excited (n$^\prime$)s$_{1/2}$ Rydberg states are denoted by subscripts $\mathrm{l}$, $\mathrm{i}$ and $\mathrm{R}$, respectively. The natural linewidths, $\Gamma$, of the states are calculated from the calculated lifetime data of Refs.~\cite{Sahoo2008, IskrenovaTchoukova2008}. The wavelength, $\lambda$, and frequency, $\omega$, are taken from literature\cite{Karlsson1999}, with $\Delta \omega$ being the energy difference from the intermediate state being on resonance. The process is non-linear in power, $P$, where the power is distributed over an area, $A$. The oscillator strengths, $f$, are calculated using the matrix dipole moments from the WSKB calculations above. The Lorentz radius is $r_0=e^2/4\pi \epsilon_0 m c^2$. The Clebsch-Gordan coefficients e.g. $\left\langle J_{\mathrm{R}} 1 m_{\mathrm{R}}-q \mid J_\mathrm{i} m_\mathrm{i}\right\rangle$ include the index $q$ for polarization ($q$=+1,0,-1 for $\sigma^+$, $\pi$, $\sigma^-$ polarizations, respectively), for the transitions of this work, linearly polarized light has the highest probability of two-photon transitions therefore $q$=0 is used herin. 
The 8p$_{1/2}$ level has a fortunate level energy compared to the 8p$_{3/2}$ level, in which an equivalent excitation path would put the two-photon transition over the ionization potential of Ba$^+$ (of 80686.30(10) cm$^{-1}$ \cite{Karlsson1999}).
The results of the one- and two-photon transition strengths are displayed in Table~\ref{tab_As}.

\begin{table}[!ht]
\caption{\label{tab_As} One-photon transitions pertinent to Rydberg state excitation in the Barium ion and quantum computing operations are compared to the enhanced two-photon transition probability P$^{2\gamma}$ for near-resonant transitions involving the intermediate 8p$_{1/2}$ state, as illustrated in Fig.~\ref{fig_Rydberg_ion_crystal}. }
\begin{ruledtabular}
\begin{tabular}{@{}cccc@{}}
One-photon transitions               & $\lambda$ (nm) & A$_{ij}$ (s$^{-1}$)  & Ref.      \\  \hline 
6s$_{1/2}$ $\rightarrow$ 6p$_{3/2}$  & 455.4          & 1.11$\times$10$^8$ & \cite{Karlsson1999}      \\
6s$_{1/2}$ $\rightarrow$ 5d$_{5/2}$  & 1762.17          & 2.5$\times$10$^5 \ll$                 &    \cite{Yum2017}  \\
5d$_{5/2}$ $\rightarrow$ 6p$_{3/2}$  & 614.17         & 4.12$\times$10$^7$ & \cite{Karlsson1999}      \\
6p$_{3/2}$ $\rightarrow$ 7s$_{1/2}$  & 489.99         & 1.04$\times$10$^8$ & \cite{Karlsson1999}      \\
7s$_{1/2}$ $\rightarrow$ 8p$_{1/2}$  & 526.75            & 1.3$\times$10$^5$  & \cite{Barakhshan2022} \\
8p$_{1/2}$ $\rightarrow$ 36s$_{1/2}$ & 528.28           & 5.2$\times$10$^3$ &           \\
8p$_{1/2}$ $\rightarrow$ 37s$_{1/2}$ & 527.60            & 4.4$\times$10$^3$  &           \\
8p$_{1/2}$ $\rightarrow$ 38s$_{1/2}$ & 526.97          & 3.8$\times$10$^3$  &           \\
8p$_{1/2}$ $\rightarrow$ 39s$_{1/2}$ & 526.40           & 3.2$\times$10$^3$ &            \\
8p$_{1/2}$ $\rightarrow$ 40s$_{1/2}$ & 525.88        & 2.8$\times$10$^3$  &           \\
8p$_{1/2}$ $\rightarrow$ 50s$_{1/2}$ & 522.38           & 7.3$\times$10$^2$ &           \\
8p$_{1/2}$ $\rightarrow$ 60s$_{1/2}$ & 520.59            & 2.6$\times$10$^2$  &           \\
 \hline 
Two-photon transitions$^\dagger$               &                &              \footnotesize{1 mW/cm$^2$}   &      \footnotesize{100 mW/cm$^2$}      \\
               &                &                P$_{2\gamma}$ (s$^{-1}$)    &       P$_{2\gamma}$ (s$^{-1}$)     \\ \hline 
7s$_{1/2}$ $\rightarrow$ 36s$_{1/2}$ & 527.52           &       1.9$\times$10$^4$      &     1.9$\times$10$^8$   \\
7s$_{1/2}$ $\rightarrow$ 37s$_{1/2}$ & 527.17           &      6.5$\times$10$^4$ &  6.5$\times$10$^8$  \\
7s$_{1/2}$ $\rightarrow$ 38s$_{1/2}$ & 526.86         &      8.6$\times$10$^5$  &  8.6$\times$10$^9$ \\
7s$_{1/2}$ $\rightarrow$ 39s$_{1/2}$ & 526.57          &    3.3$\times$10$^5$     &  3.3$\times$10$^9$  \\
7s$_{1/2}$ $\rightarrow$ 40s$_{1/2}$ & 526.31            &     4.5$\times$10$^4$  &  4.5$\times$10$^8$\\
7s$_{1/2}$ $\rightarrow$ 50s$_{1/2}$ & 524.56           &      2.7$\times$10$^3$    &  2.7$\times$10$^7$ \\
7s$_{1/2}$ $\rightarrow$ 60s$_{1/2}$ & 523.65           &     3.0$\times$10$^2$   &   3.0$\times$10$^6$  \\ 
\end{tabular}
\end{ruledtabular}
\footnotesize{
$\dagger$ - Two-photon transition rates calculated to be highest using linearly polarized light and $|$m$_f$$|$=1/2 levels}
\end{table}

\section{Methods: Dynamic polarizabilities} 

\begin{figure}[t]
\includegraphics[width=9cm]{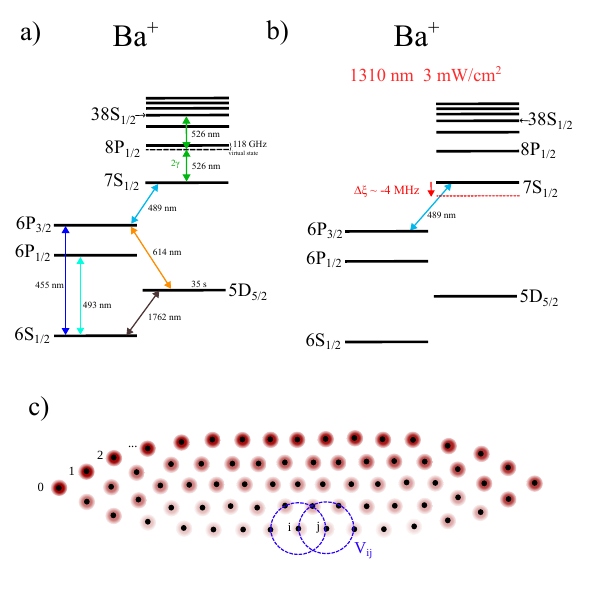}
\caption{\label{fig_Rydberg_ion_crystal} a) Schematic of relevant energy levels and transitions in the Barium ion, the two-photon transition to Rydberg energy levels is indicated at 526~nm. b) Schematic showing the large energy shift of the 7s$_{1/2}$ level with 1310-nm AC-Stark shifting light. c) Schematic of varying spot power intensity of AC-Stark shifting 1310-nm light, overlayed with a 2d Wigner ion crystal, used to create unique addressing frequencies for the 7s$_{1/2}$ $\rightarrow$ 38s$_{1/2}$ transition. }
\end{figure}

\begin{figure}[t]
\includegraphics[width=9cm]{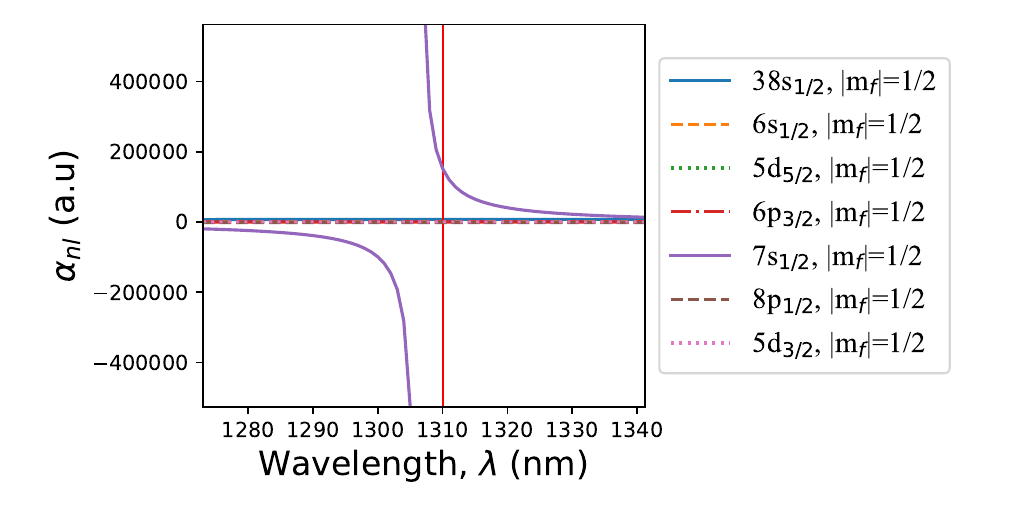}
\caption{\label{fig_1310} Polarizability, $\alpha_{n^{\prime}l^{\prime}}$, of Ba$^+$ indicating the large polarizability of the 7s$_{1/2}$ state around 1310~nm relative to lower-lying and Rydberg states.}
\end{figure}

The dipole matrix elements calculated above allow determination of the dynamic polarizability, $\alpha_{nl}(\omega)$, and through it the AC-stark energy shifts (i.e. light shift), $\Delta \xi_v(\omega)$, expected for each state due to a laser field with electric field strength $E$, as

\begin{equation}
    \Delta \xi_v(\omega)=-\frac{1}{2} \alpha_{n^{\prime}l^{\prime}}(\omega) E^2 .
    \label{eqn_STARK}
\end{equation}

\begin{table*}[!ht]
\caption{\label{tab_polar} Calculated dynamic polarizabilities $\alpha_{n^{\prime}l^{\prime}}$ for states of interest in Ba$^+$, at wavelengths of 532~nm and 1310~nm. Corresponding AC-stark energy shifts for 1310~nm light at two intensities is shown.}
\begin{ruledtabular}
\begin{tabular}{@{}cccccc@{}}
            &                    &        @ 532 nm           &     @ 1310 nm     & 1310 @ 3 mW /cm$^2$           & 1310 nm @ 30 mW /cm$^2$      \\
State & Energy level (cm$^{-1}$) & $\alpha_{nl}$ (a.u.)  & $\alpha_{nl}$ (a.u.) & $\Delta \xi_R$ (MHz) & $\Delta \xi_R$ (MHz) \\
\hline
6s$_{1/2}$  & 0                  & 5.7$\times$10$^2$  & 1.4$\times$10$^2$   & -0.0010(1)               & -0.010(1)            \\
5d$_{3/2}$  & 4873.852           & -1.9$\times$10$^2$ & 3.2$\times$10$^2$   & 0.00229(23)              & -0.0229(23)          \\
5d$_{5/2}$  & 5674.807           & -6.0$\times$10$^2$ & 2.2$\times$10$^2$   & 0.00158(16)              & -0.0158(16)          \\
6p$_{3/2}$  & 21952.404 & 7.9$\times$10$^2$  & 6     & -4.3(4)$\times$10$^{-5}$ & -0.00043(4)          \\
7s$_{1/2}$  & 42355.175          & -6.0$\times$10$^2$ & $^\dagger$1.6-6$\times$10$^5$ & -1.15(13) to -4.3(4)     & -11.5(1.1) to -43(4) \\
8p$_{1/2}$  & 61339.5   & -3.0$\times$10$^2$ & -1.6$\times$10$^3$  & 0.0115(11)               & 0.115(11)            \\
38s$_{1/2}$ & 80315.93           & 1.2$\times$10$^2$  & 7.3$\times$10$^3$   & -0.052(5)                & -0.52(5)            
\end{tabular}
\end{ruledtabular}
\footnotesize{
$\dagger$-Large gain in polarizability could be obtained by optimising the 1310~nm light-shifting laser frequency}
\end{table*}

There are multiple contributions to the total $\alpha_{n^{\prime}l^{\prime}}(\omega)$, here we adopt the approach of Ref.~\cite{Bhowmik2024}, where 

\begin{equation}
    \alpha_{n^{\prime}l^{\prime}}(\omega)=\alpha_{n^{\prime}l^{\prime}}^C(\omega)+\alpha_{n^{\prime}l^{\prime}}^{V C}(\omega)+\alpha_{n^{\prime}l^{\prime}}^V(\omega).
    \label{eqn_dyn_polar}
\end{equation}

The core polarizability, $\alpha_{n^{\prime}l^{\prime}}^C(\omega)$, is calculated in the literature\cite{Das2022} to be $\alpha_{n^{\prime}l^{\prime}}^C(\omega=0)$~=~10.79~a.u. which varies negligibly with frequency in comparison to the valance polarizability $\alpha_{n^{\prime}l^{\prime}}^V(\omega)$. The contribution of the valence electron to the core polarizability $\alpha_{n^{\prime}l^{\prime}}^{V C}(\omega)$ is taken as a negligibly small contribution due to the relatively tightly bound core electrons i.e. $\alpha_{n^{\prime}l^{\prime}}^{V C}(\omega) \approx 0$.

The valance polarizability, $\alpha_{n^{\prime}l^{\prime}, J_{n^{\prime}l^{\prime}}}^V(\omega)$, can be calculated as

\begin{equation}
    \begin{array}{r}
\alpha_v^V(\omega)=\alpha_{n^{\prime}l^{\prime}, J_{n^{\prime}l^{\prime}}}^{(0)}(\omega)+ \\
\frac{3 \cos ^2 \theta-1}{2} \frac{3 M_J^2-J(J+1)}{J(2 J-1)} \alpha_{v, J}^{(2)}(\omega)
\end{array}
    \label{eqn_Valence_pol}
\end{equation}

where $\alpha_{n^{\prime}l^{\prime}, J_{n^{\prime}l^{\prime}}}^{(0)}(\omega)$ is the scalar polarizability contribution given by 

\begin{equation}
    \begin{array}{r}
    \alpha_{n^{\prime}l^{\prime}, J_{n^{\prime}l^{\prime}}}^{(0)}(\omega)=\frac{2}{3(2 J+1)} \sum_{nl} d_{nl}^{n^{\prime}l^{\prime}},
    \end{array}
    \label{eqn_scalar_pol}
\end{equation}

and $\alpha_{n^{\prime}l^{\prime}, J_{n^{\prime}l^{\prime}}}^{(2)}$ is the tensor polarizability contribution, given by

\begin{equation}
\begin{aligned}
\alpha_{v, J}^{(2)}(\omega)= & 4 \sqrt{\frac{5 J(2 J-1)}{6(J+1)(2 J+1)(2 J+3)}} \times \\
& \sum_n(-1)^{J_n+J}\left\{\begin{array}{ccc}
J & 1 & J_n \\
1 & J & 2
\end{array}\right\} d_{nl}^{n^{\prime}l^{\prime}} .
\end{aligned}
    \label{eqn_tensor_pol}
\end{equation}

For both Equation~\ref{eqn_scalar_pol} and Eqn.~\ref{eqn_tensor_pol}, $d_{nl}^{n^{\prime}l^{\prime}}$ can be determined using the dipole element calculated using Eqn.~\ref{eqn_Bnl} via

\begin{equation}
    \begin{array}{r}
    d_{nl}^{n^{\prime}l^{\prime}}=\frac{e^2 |R_{n l}^{n^{\prime} l^{\prime}}|^2 \omega_{n v}}{\omega_{n v}^2-\omega^2}
    \end{array}
    \label{eqn_dnv}
\end{equation}

The equations above do not consider the calculation at the hyperfine level, this can be found in Ref.~\cite{Bhowmik2024}. Table.~\ref{tab_polar} shows the results of these calculations at two particular wavelengths of interest, 532~nm and 1310~nm.

The 532~nm wavelength corresponds to a Nd:YAG laser wavelength commonly used to drive Raman transitions for gates in trapped Ba$^+$. As can be seen in Table.~\ref{tab_polar}, at 532~nm, all of the states of interest have similar magnitudes for polarization, moreover they would require Watts of laser power to achieve the desired Stark shifts, see e.g. Ref.~\cite{Staanum2002}.

However, as Fig.~\ref{fig_1310} shows, the 7s$_{1/2}$ state coupled with 1310-nm light (which is readily available and with very low noise due to its use in the `O-band' for telecommunications \cite{Morton2018}) has a very high polarizability compared to other states at this wavelength (five orders of magnitude greater than the ground 7s$_{1/2}$ state) which is further shown in Table.~\ref{tab_polar}. Note that 526~nm and 489~nm are also readily available diode wavelengths \cite{Miyoshi2010,Knig2019}. For clarity only the $|m_f|$=1/2 sub-states are plotted; there is little difference in polarizability for higher $|m_f|$=3/2, 5/2 states far away from the resonant transition wavelengths.

\begin{figure}[t]
\includegraphics[width=8.5cm]{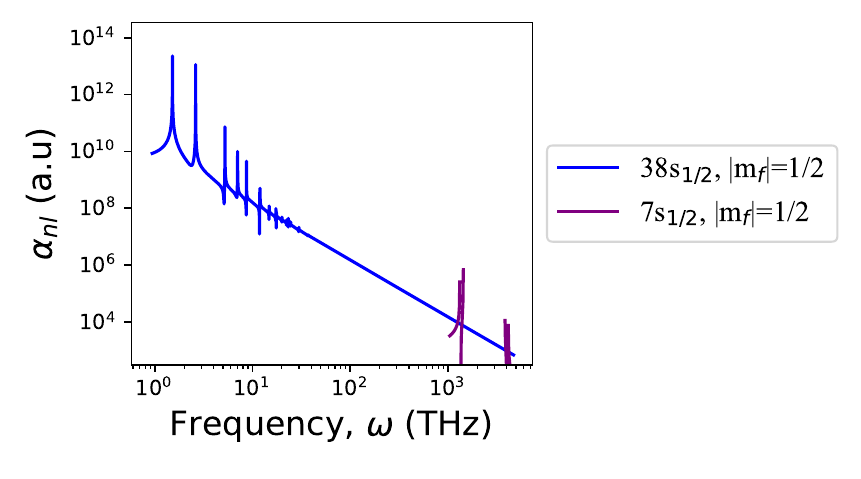}
\caption{\label{fig_rydberg_freq} Trend of the polarizability, $\alpha_{n^{\prime}l^{\prime}}$, of (n$^\prime$)s$_{1/2}$ Rydberg states in Ba$^+$ (e.g. 38s$_{1/2}$), the polarizability for 7s$_{1/2}$ is shown for comparison.}
\end{figure}

In this wavelength region the dynamic polarizability of the Rydberg states has a low magnitude and varies monotonically, similar to the lower-lying states, as the range is off-resonant from any transitions to the 38s$_{1/2}$ state. This can be seen in Figure~\ref{fig_rydberg_freq}, where a high polarizability is found for decreasing frequency. It is predicted that the polarizability takes the form of a series of poles for Rydberg states.
This is a result of the density of states below and above being roughly equal, therefore their contributions cancel and the most important contributions to polarizability come instead from nearby Rydberg states, see Ref.~\cite{Yerokhin2016}.

This effect is apparent in Fig.~\ref{fig_rydberg_freq}, as the polarizability changes from having a negative component until the frequency lowers, then only single poles are observed.

Assuming Gaussian profiles, with spacing, $W$, of 5 $\mu$m, and with 0.5 $\mu$m micromotion or laser instability, then the average field intensity experienced is given by
\begin{equation}
    E^2=\frac{2 P}{\varepsilon_0 c A} \int_{-0.25}^{0.25} \exp \left(\frac{-\left(r-r_0\right)^2}{W}\right) dr,
\end{equation}
which is used to evaluate Eqn.~\ref{eqn_STARK} to give the values in Table.~\ref{tab_polar}.

\FloatBarrier

Fig.~\ref{fig_Rydberg_ion_crystal}a), highlights a few transitions in the Ba$^+ $ ion and the possible effect of the proposed scheme on adding selectivity to the  STIRAP process.\\

\section{Discussion}

The mass-independent entanglement of Rydberg ion excitation supports scalability, enabling 2D or 3D Wigner crystals with tailored spacings for uniform entanglement. A trap can be designed to counter the increasing interaction strength ($V_{ij}$) with rising principal quantum number ($n$) by leveraging the $1/r^3$ dependence of $V_{ij}$, maintaining consistent interaction spacing ($V_u$) that may vary across the trap. This ion-light interaction is illustrated in Fig.~\ref{fig_Rydberg_ion_crystal}b, c) for a 2D Wigner crystal. Fig.~\ref{fig_Rydberg_ion_crystal}c) depicts a Wigner crystal ion lattice, reliably generated by ion trap groups with stable, reproducible ion positions for over 500 ions, as shown in Ref.~\cite{Guo2024}. The surface potential must be tuned to the equilibrium spacing for optimal Rydberg interaction strength; otherwise, interaction strengths require calculation, possibly using a camera to map ion positions, followed by calibration to minimize decoherence from varying interactions.

Fig.~\ref{fig_Rydberg_ion_crystal}c) also shows a varying light field intensity across each ion (transparent red), relevant for the interacting ion species, here assumed to be Ba$^+$ due to its experimental advantages. A diffraction grating with suitable spacing for a 1D ion chain could serve as a simpler initial demonstration. Alternatively, combining multiple spatial light modulators (SLMs) \cite{Lei2015} or using structured light optics \cite{Dorrah2025} could enable unique intensity variations at each ion in a 3D lattice.

Microwave-field dressing is needed to mix high-lying (n$'$)$s_{1/2}$ states with the nearest (n$'$)$p_{1/2}$ states, canceling Rydberg state polarizability and reducing susceptibility to the trapping RF field \cite{Booth2018}. Additionally, microwave dressing enables tuning of interaction strength or switching it on/off, a feature absent in neutral atoms reliant on the weaker, constant van der Waals interaction, which is insufficient for trapped ion Rydberg states \cite{Giudici2024}. This opens new possibilities for simulations using Rydberg dipolar interactions in trapped ions \cite{Mokhberi2020}.

Initially, adding two wavelengths (526 nm, 489 nm) to the Ba$^+$ excitation scheme might appear disadvantageous due to potential decoherence and added optics. However, these transitions eliminate the need for UV light and enhance and allow selectivity through the AC-Stark shift of the 7s$_{1/2}$ level from the intensity of the 1310-nm light. Similar approaches may work for other alkali-like ions (Mg$^+$, Ca$^+$, Sr$^+$, ...), but Ba$^+$, aside from Ra$^+$ (lacking stable isotopes), has the lowest ionization potential for excitation to the Rydberg state without UV light. As shown in Fig.~\ref{fig_1310} and Tab.~\ref{tab_polar}, the 7s$_{1/2}$ dynamic polarizability is highly sensitive to the frequency of the 1310-nm light.  While low-noise 1310-nm laser infrastructure exists for telecommunications, additional optics for phase or power stabilization, or filtering via optical cavities, may be required due to this sensitivity. Experimental validation is essential in all cases.

SLMs are well-suited for this architecture due to their commercial refresh rates in the kHz range, pixel densities in the thousands by thousands, pixel spacings on the order of micrometers, and $>$90$\%$ reflection at 1310 nm. Phase masks for intensity mapping onto equilateral triangle spacings have been demonstrated \cite{PinheirodaSilva2022}. The high refresh rate enables rapid compensation for ion position drift when paired with a CCD or an array of orthogonal CCDs, while the micrometer-scale pixel spacing and high density facilitate straightforward optical addressing of the Wigner crystal.
Raman transitions for qubit operations are highly sensitive to intensity variations or cross-talk between addressed ions. In contrast, using the AC-Stark shift induced by 1310-nm light, spatially modulated by a SLM, enables compensation for such variations through the laser bandwidth of the 6p$_{3/2}$ $\rightarrow$ 7s$_{1/2}$ (489 nm) transition.

\section{Conclusion}

The proposed method enables UV-free, individually addressable entanglement of Barium ion qubits using Rydberg states and AC-Stark-shift selectivity, towards tackling scalability issues in ion-trap quantum computing independent of ion number.
A two-photon transition at 526 nm eliminates the need for UV light, reducing risks to surface traps and simplifying optics. The high polarizability of the 7$s_{1/2}$ state at 1310 nm allows significant AC-Stark shifts with mW-range laser power for MHz-level shifts, compared to W-range power needed in prior proposals using less polarizable states \cite{Staanum2002}. 
The selectivity can then be introduced by tuning of the 7s$_{1/2}$ $\rightarrow$ (n')s$_{1/2}$ transition frequency.
This supports individual addressing and sub-microsecond entangling gates via strong, microwave-tunable Rydberg dipolar interactions.
The Supersymmetric Wentzel-Kramers-Brillouin (SWKB) method was used to calculate critical dipole matrix elements and transition strengths, confirming the feasibility.
The sensitivity of the 7$s_{1/2}$ to 1310-nm light leverages existing telecommunications infrastructure for practicality. 
Experimental validation is needed to confirm performance and extend the approach to other high-fidelity alkali-like ion systems.

\FloatBarrier

\begin{acknowledgments}
\section{Acknowledgements}
The authors thank Chris Monroe and Crystal Noel at Duke University, and Weibin Li at the University of Nottingham, for valuable insight and discussions.

\end{acknowledgments}

\bibliography{export}

\FloatBarrier





\end{document}